\documentclass[useAMS,usenatbib]{mn2e}

\usepackage[pdftex]{graphicx}
\usepackage{fancyhdr}
\usepackage{hyperref}
\usepackage[dvips]{epsfig}
\usepackage{color}

\title[Radiatively-Driven Black-Hole Winds Revisited]{Radiatively-Driven Black-Hole Winds Revisited}
\author[R. Yamamoto and J. Fukue]{R. Yamamoto \thanks{E-mail:
ryoya@astro-osaka.jp} and J. Fukue \thanks{E-mail:
fukue@cc.osaka-kyoiku.ac.jp}\footnotemark[0]\\
Astronomical Institute, Osaka Kyoiku University, 
Asahigaoka, Kashiwara, Osaka 582-8582, Japan}

\begin{document}

\date{}

\pagerange{\pageref{firstpage}--\pageref{lastpage}} \pubyear{2021}

\maketitle

\label{firstpage}

\begin{abstract}
We examine general relativistic radiatively-driven spherical winds,
using the basic equations for relativistic radiation hydrodynamics
under the moment formalism.
Moment equations are often closed,
using the equilibrium diffusion approximation,
which has an acausal problem,
and furthermore, gives nodal-type critical points.
Instead,
we use the nonequilibrium diffusion approximation
with a closure relation of a variable Eddington factor, $f(\tau,\beta)$, 
where $\tau$ is the optical depth and $\beta$ is the flow speed normalized by the speed of light.
We then analyze the critical properties in detail for several parameters,
and found that there appear saddle-type critical points
as well as nodal type and spiral one.
   The most suitable type is the saddle one, which appears in a region close to a black hole.
We also calculate transonic solutions with typical parameters,
and show that the luminosity is almost comparable to the Eddington luminosity, 
the gas is quickly accelerated in the vicinity of the black hole,
   and wind terminal speeds are on the order of 0.1--0.3~$c$.
These results of radiatively-driven black hole winds
can be applied, e.g., to ultra-fast outflows (UFOs), 
   which are supposed to be fast outflows from the vicinity of super massive black holes.
\end{abstract}


\section{Introduction}

Supermassive black holes (SMBHs) in active galactic nuclei (AGNs) are known to have luminous accretion discs 
(Shakura \& Sunyeav 1973; Kato et al. 2008 and reference therein).
Various evidences of mass outflows from a luminous disc are frequently observed, but their origins and dynamics are not yet well clarified.
For example, 10-15\% of quasars are classified to BAL (broad absorption line) quasars, which show broad, blue-shifted and strong absorption lines 
in the spectra of higher-order ionized atoms such as 
N{\footnotesize V},  C{\footnotesize I\hspace{-1pt}V}, and Si{\footnotesize I\hspace{-1pt}V},
due to optical or UV observations 
(Weymann et al. 1991; Hamann et al. 1993; Gibson et al. 2009; Allen et al. 2011).
From the center of BAL quasars, it is supposed that
accretion disc winds of $10000-30000 ~\rm km~s^{-1}$ (0.01--$0.1c$) are blowing off from the black hole vicinity.
The column density of these outflows is estimated to be $10^{23}$--$10^{24}~\rm cm^{-2}$ 

The existence of high speed winds is also found by X-ray observation.
In about 40\% of AGNs, the absorption lines of Fe{\footnotesize XXV} and Fe{\footnotesize XXI\hspace{-1pt}V} are found in the outflow.
Their corresponding velocities are 0.1--$0.3c$ and these are called ultra-fast outflows (UFOs) 
{(e.g., Tombesi et al. 2010, 2011, 2012, 2013, 2014)}.
The column density is estimated to be $10^{22}$--$10^{23}\rm cm^{-2}$.
UFOs are thought to have energies comparable to energetic jets, 
which may have a significant impact on AGN feedbacks such as star formation and SMBH growth in the bulge
{(e.g., Nayakshin 2010; Wagner et al. 2013; King \& Muldrew 2016;
Longinotti 2018)}.

Theoretically, spherically symmetric optically-thick winds 
driven by radiation pressure under general relativity
have been investigated by several researchers
(Lindquist 1966; Castor 1972; Cassinelli \& Hartmann 1975; 
Ruggles \& Bath 1979; Mihalas 1980;
Quinn \& Paczy\'{n}ski 1985; Paczy\'{n}ski 1986, 1990; 
Paczy\'{n}ski \& Pr\'{o}szy\'{n}ski 1986;
Turolla et al. 1986; Nobili et al. 1994; 
Akizuki \& Fukue 2008, 2009),
using radiation hydrodynamical equations under the moment formalism
(Thorne 1981; Park 2006; Takahashi 2007; see also Kato \& Fukue 2020).
Many of these studies adopted the equilibrium diffusion approximation,
where the radiation temperature is equal to the gas one,
in order to close the moment equations.
However,
the usage of the equilibrium diffusion approximation
in the moving media is physically questionable,
since it permits thermal pulses to travel faster than the speed of light,
and notoriously acausal, as was stated by Thorne et al. (1981).
Furthermore,
under the equilibrium diffusion approximation,
the transonic points are proved to be always nodal 
in the nonrelativistic regime (Fukue 2014).

Instead of the equilibrium diffusion approximation,
several studies adopted
the nonequilibrium diffusion approximation,
where the radiation temperature is not equal to the gas one,
or the radiation pressure is not expressed by the gas temperature
(Nobili et al. 1994; Akizuki \& Fukue 2008, 2009).
In this case,
some closure relation is necessary,
and the Eddington approximation is usually adopted.
However,
a simple Eddington approximation with the Eddington factor of 1/3
in the nonrelativistic regime
 is known to bring a pathological behavior in the relativistic regime
(e.g., Turolla \& Nobili 1988; Nobili et al. 1991;
Turolla et al. 1995; Dullemond 1999; Fukue 2005).
Namely, the moment equations under the simple Eddington approximation
in the relativistic regime have singular points,
which are purely mathematical artifacts of the moment expansion
(Dullemond 1999).
Hence,
instead of the simple Eddington approximation,
the variable Eddington factor has been used
(Noboli et al. 1994; Akizuki \& Fukue 2008, 2009).
In Akizuki and Fukue (2008, 2009), for example,
the velocity-dependent variable Eddington factor was used.
In their studies, however,
the gas pressure was dropped for simplicity.

Thus, in this paper, 
in order to resolve the transonic black hole winds again,
we investigate the general relativistic radiation hydrodynamical winds,
under the nonequilibrium diffusion approximation
with the help of a variable Eddington factor $f(\tau,\beta)$,
which depends both on the optical depth $\tau$ and the flow speed $v$
($=\beta c$)
(Akizuki \& Fukue 2008, 2009),
and examine the topological nature of the critical points
of the black hole winds in detail.

In the next section we describe the basic equations 
for general relativistic winds driven by radiation pressure 
in the spherical symmetric case.
In section 3 we derive tne wind equations from the basic equations.
In section 4 we show the loci of critical points,
and examine their type,
while in section 5 we solve and show the transonic solutions 
with typical parameters, respectively.
The final section is devoted to concluding remarks.


\section{Basic equations}

In this section, 
we describe the basic equations for the present spherically symmetric,
optically-thick, steady wind driven by radiation pressure 
from the vicinity of the central black hole of mass $M$. 
General relativistic radiation hydrodynamical equations
have been derived by several studies
(Lindquist 1966; Anderson \& Spiegel 1972; Thorne 1981;
Udey \& Israel 1982; Nobili et al. 1993; {Park 1993, 2006}; Takahashi 2007;
see also Kato et al. 2008; Kato and Fukue 2020).

For gas, the continuity equation is
\begin{equation} \label{eq-conti}
    4\pi r^2 \rho c u =\dot{M}, 
\end{equation}
where $\rho$ is the proper gas density, $c$ the speed of light,
$u$ the radial component of the four velocity, 
and $\dot{M}$ the constant mass-loss rate.
Using the proper three velocity $v$ and $\beta$ ($\equiv v/c$),
the four velocity $u$ is expressed as
$u=y\beta$,
where
$y=\gamma\sqrt{g_{00}}$, $\gamma=1/\sqrt{1-\beta^2}$, $g_{00}=1-r_{\rm S}/r$,
$r_{\rm S}$ being the Schwartzschild radius ($r_{\rm S}=2GM/c^2$),

The equation of motion is
\begin{equation} \label{eq-motion}
    u\frac{d u}{dr}+\frac{r_{\rm S}}{2r^2}+\frac{y^2}{\varepsilon +p}\frac{d p}{dr} =\frac{y}{\varepsilon +p}\frac{\rho \overline{\kappa}_{\rm F}}{c} F_{0},
 \end{equation}
where $p$ is the gas pressure, $\overline{\kappa}_{\rm F}$
($=\kappa+\sigma$)
the frequency-integrated flux-mean opacity 
for absorption $\kappa$ and scattering $\sigma$,
and $F_0$ the radiative flux in the comoving frame.
The Lorentz transformation of the radiation moments 
in the fiducial observer frame
and those in the comoving one is expressed as 
\begin{equation}
    F_0=\gamma^2[(1+\beta^2)F-\beta(cE+cP)],
\end{equation}
where $E$ is the radiation energy density, $F$ the radial component of the radiative flux, and $P$ the $rr$ component of the radiation stress tensor
in the fixed frame.
In contrast to Akizuki and Fukue (2009),
in the present study
we include the gas pressure term,
which creates the transonic points in the flow.

Instead of the radiative equilibrium in Akizuki and Fukue (2009),
we use the full form of the energy equation for gas: 
\begin{equation} \label{eq-energy}
    \frac{c}{r^2}\frac{d}{dr}[ r^2(\varepsilon -\rho c^2)u]+c\frac{p}{r^2}\frac{d}{dr}(r^2 u)=-\rho (j_{0}-\overline{\kappa}_{\rm E}cE_{0} ),
\end{equation}
where  $\varepsilon$ is the gas internal energy including the rest-mass energy, 
and expressed as
\begin{equation} \label{EoS}
    \varepsilon=\rho c^2 +\frac{p}{\Gamma-1},
\end{equation}
$\Gamma$ being the ratio of specific heats.
Furthermore,
$\overline{\kappa}_{\rm E}$ ($=\kappa$)
is the frequency-integrated energy-mean absorption opacity.
Moreover,
$E_0$ is the radiation energy density in the comoving frame,
and the Lorentz transformation is 
\begin{equation}
    cE_0=\gamma^2[cE-2\beta F+\beta^2cP].
\end{equation}
In addition, $j_0$ is the frequency-integrated emissivity, 
and can be written 
under the local thermodynamic equilibrium (LTE) condition as
\begin{equation} \label{LTE}
    j_0=4\pi \overline{\kappa}_E B(T),
\end{equation}
where $B$ is the frequency-integrated blackbody intensity,
$B(T)=\sigma_{\rm SB}T^4/\pi$,
$\sigma_{\rm SB}$ being Stephan-Boltzmann constant,
and $T$ the gas temperature.

The equation of state is
\begin{equation}
    p=\rho \frac{\mathcal{R}}{\mu} T,
\end{equation}
where $\mathcal{R}$ is the gas constant, 
$\mu$ ($=0.5$, fully ionized hydrogen plasma) the mean molecular weight.
We assume that the black hole winds are sufficiently hot,
and the gas is fully ionized.

The adiabatic sound speed, $c_{\rm s}$, is defined as
\begin{equation}
    c_{s}^2=c^2\left(\frac{dp}{d\varepsilon}\right)_{\rm adiabatic}=c^2\frac{\Gamma p}{\varepsilon+p}.
\end{equation}

For radiation, the 0-th moment equation is
\begin{equation}\label{0th-moment}
    \frac{d}{dr}(4 \pi r^2 g_{00}F)=4\pi r^2 \rho y\left(j_0-\overline{\kappa}_{\rm E}cE_0-\beta \overline{\kappa}_{\rm F} F_{0}\right), 
\end{equation}
while the first moment equation is
\begin{eqnarray} \label{1st-moment}
    \frac{d}{dr}(4 \pi r^2 g_{00}cP)&=&4\pi r\left(1-\frac{3r_{\rm S}}{2r}\right)(cE-cP)\nonumber \\
    &&-4\pi r^2 \rho y\bigl[\overline{\kappa}_{\rm F} F_{0}
    -\beta\left(j_0-\overline{\kappa}_{\rm E}cE_0\right)\bigr].\nonumber \\
    &&
\end{eqnarray}
The $rr$ component of the radiation stress tensor, $P_0$, 
in the comoving frame is written as 
\begin{equation}
    cP_0=\gamma^2[\beta^2cE-2\beta F+cP].
\end{equation}

In the present study, 
we do not assume the equilibrium diffusion approximation, 
where $P_0$ is expressed in terms of the gas temperature,
as $P_0 = aT^4/3$,
but assume the nonequilibrium diffusion approximation,
where
$P_0$ or the radiation temperature is an independent variable,
and some closure relation is necessary.

As a closure relation, we adopt the Eddington approximation:
\begin{equation}\label{edd-app}
    P_0=f(\beta,\tau)E_0,
\end{equation}
where $f(\beta,\tau)$ is the variable Eddington factor 
which depends both on the optical depth and the flow speed
(Tamazawa et al. 1975; Abramowicz et al. 1991; Akizuki \& Fukue 2008, 2009): 
\begin{equation}
    f(\beta,\tau)=\frac{\gamma(1+\beta)+\tau}{\gamma(1+\beta)+3\tau}
\end{equation}

Finally,
we introduce the optical depth variable $\tau$
by 
\begin{eqnarray} \label{def-tau}
    d\tau &=& -\rho\overline{\kappa}_{\rm F}\gamma(1-\beta \cos\theta)\sqrt{g_{11}} dr \nonumber \\
          &=& -\rho(\kappa+\sigma)\gamma(1-\beta) \frac{dr}{\sqrt{g_{00}}} ,
\end{eqnarray}
where $\theta$ is the angle between the velocity and the line-of-sight,
and set as $\theta=0$ in the present wind case
(cf. Abramowicz et al. 1991; Nied\'{z}wiecki \& Zdziarski 2006; Fukue 2011).

Using equations (\ref{eq-conti}), (\ref{eq-motion}), (\ref{0th-moment}), and (\ref{1st-moment}), 
we derive the additional equation, the Bernoulli equation, for the present case:
\begin{equation} \label{bel-0}
    \dot{M}\frac{\varepsilon +p}{\rho}y+4\pi r^2 g_{00}F =\dot{E} ,
\end{equation}
where $\dot{E}$ is a Bernoulli constant.
In the relativistic regime, where the flow velocity becomes large in comparison with the speed of light, 
there appear advection terms in this Bernoulli equation,
when the radiative flux $F$ in the inertial frame
is converted to that $F_0$ in the comoving one.

Since we treat the spherically symmetric flow,
instead of the linear flux $F$ and radiation pressure $P$,
we use the spherical variables $L$ and $Q$ defined by
\begin{eqnarray}
   L &\equiv& 4\pi r^2 g_{00} F,
\\
   Q &\equiv& 4\pi r^2 g_{00} cP,
\end{eqnarray}
where $L$ is the luminosity measured by an observer at infinity.
Moreover, with the black hole winds in mind, 
we define and use nondimensional variables:
\begin{equation} \label{non-d-variable}
    \hat{r}\equiv \frac{r}{r_{\rm S}},~~~~\beta \equiv \frac{v}{c},~~~~\alpha_{\rm s} \equiv \frac{c_{\rm s}}{c},~~~~\hat{L} \equiv \frac{L}{L_{\rm E}},~~~~\hat{Q} \equiv \frac{Q}{L_{\rm E}},
\end{equation}
and nondimensional parameters:
\begin{equation} \label{non-d-parameter}
    m\equiv\frac{M}{M_{\odot}},~~~~\dot{m}\equiv\frac{\dot{M}c^2}{L_{\rm E}},~~~~\dot{e}\equiv\frac{\dot{E}}{L_{\rm E}},
\end{equation}
where $L_{\rm E}$ ($\equiv 4\pi cGM/\overline{\kappa}_{\rm F})$ is the Eddington luminosity,
and $\dot{M}_{\rm E}$ ($\equiv {L_{\rm E}}/{c^2})$ the critical mass-loss rate.

In the previous paper of the UFOs observation, they estimated UFOs mean mass-loss rate is $\dot{M} \sim 0.01-1\rm M_{\odot} yr^{-1}$ (Tombesi et al. 2012).
Specifically, the quaser PG1211+143 has a SMBH with a mass of $\sim 10^8 \rm M_\odot$ at its center, 
and using the nondimensional mass-loss rate, it is estimated $\dot{m}\sim$ 1--$10^2$.

\section{Wind equations}

Eliminating the gas-pressure gradient from equations (\ref{eq-motion}) and (\ref{eq-energy}), 
and the density using equation (\ref{eq-conti}), 
we obtain the wind equation on the velocity:
\begin{eqnarray}\label{wind-beta}
        \frac{d\beta}{dr}&=&\frac{1}{y^2(\beta^2-\alpha ^2_{\rm s})}\Biggl\{\beta\Biggl[-\frac{r_{\rm S}}{2r^2}\nonumber\\
        &&+\frac{2}{r}\left(1-\frac{3r_{\rm S}}{4r}\right)\alpha ^2_{\rm s}+\frac{\delta^2y}{\gamma^2c^2}\overline{\kappa}_{\rm F}\frac{F_0}{c}\Biggr] \nonumber \\
        &&+\frac{\delta^2y}{\gamma^2c^2}\frac{\Gamma-1}{c}\left(j_0-c\overline{\kappa}_{\rm E}E_0\right)\Biggr\} ,
\end{eqnarray}
where
\begin{equation}
    \delta^2\equiv \frac{\rho c^2}{\varepsilon +p}=1-\frac{\alpha^2_{\rm s}}{\Gamma-1}.
\end{equation}
On the other hand, 
the gas-pressure gradient is expressed
in terms of the density and adiabatic sound as 
\begin{equation} \label{pressure-grad}
    \frac{dp}{dr}=\frac{\varepsilon+P}{\Gamma}\left(\frac{\alpha_{\rm s}^2}{\rho}\frac{d\rho}{dr}+\frac{\Gamma-1}{\Gamma-1-\alpha_{\rm s}^2}\frac{d\alpha_{\rm s}^2}{dr}\right) .
\end{equation}
Hence, eliminating the gas-pressure gradient from equation (\ref{eq-motion}) and (\ref{pressure-grad}), and the density using equations (\ref{eq-conti}), 
we obtain the wind equation on the adiabatic sound speed
\footnote{
    If we take the nonrelativistic limit ($g_{00}\rightarrow1$, $\gamma\rightarrow 1$, $\delta\rightarrow 1$)
    in the wind equations (\ref{wind-beta}) and (\ref{wind-alpha}),
    those coincide with the wind equations derived by Fukue (2014).
}:
\begin{eqnarray} 
    \frac{d\alpha^2_{\rm s}}{dr}&=&-\frac{(\Gamma-1)\delta^2}{y^2(\beta^2-\alpha^2_{\rm s})}\Biggl[\alpha ^2_{\rm s}\Biggl(-\frac{r_{\rm S}}{2r^2}+\frac{2}{r}y^2\beta^2+\frac{\delta^2y}{c^2}\overline{\kappa}_{\rm F}\frac{F_0}{c}\Biggr) \nonumber \\
    &&+\frac{\delta^2y}{c^2}\frac{\Gamma \beta^2-\alpha^2_{\rm s}}{c\beta}\left(j_0-c\overline{\kappa}_{\rm E}E_0\right)\Biggr] . \nonumber
\end{eqnarray}
\begin{equation} \label{wind-alpha}
    ~
\end{equation}

As was stated, we assume the LTE condition (\ref{LTE}) in this paper.
Then, equations (\ref{wind-beta}), (\ref{wind-alpha}), (\ref{0th-moment}), (\ref{1st-moment}), (\ref{def-tau}), and (\ref{bel-0})
are respectively rewritten as follows:
\begin{eqnarray}\label{wind-beta-re}
        \frac{d\beta}{dr}&=
        &\frac{1}{y^2(\beta^2-\alpha ^2_{\rm s})}\Biggr\{\beta\Biggl[-\frac{r_{\rm S}}{2r^2}+\frac{2}{r}\left(1-\frac{3r_{\rm S}}{4r}\right)\alpha ^2_{\rm s}\nonumber \\
        &&+\frac{\delta^2y}{4\pi r^2g_{00}\gamma^2}\frac{\overline{\kappa}_{F}}{c^3}\frac{(f+\beta^2 )L-(1+f)\beta Q}{f-\beta ^2}\Biggr]\nonumber \\
        &&+\frac{\delta^2y}{4\pi r^2g_{00}\gamma^2}(\Gamma-1)\frac{\overline{\kappa}_E}{c^3}\nonumber \\
        &&\times\left[16\pi^2g_{00} r^2B-\frac{(1+\beta^2)Q-2\beta L}{f-\beta^2}\right]\Biggr\},
\end{eqnarray}

\begin{eqnarray} \label{wind-alpha-re}
        \frac{d\alpha^2_{\rm s}}{dr}&=
        &-\frac{(\Gamma-1)\delta^2}{y^2(\beta^2-\alpha^2_{\rm s})}\Biggl\{\alpha ^2_{\rm s}\Biggl[-\frac{r_{\rm S}}{2r^2}+\frac{2}{r}y^2\beta^2\nonumber\\
        &&+\frac{\delta^2y}{4\pi r^2g_{00}}\frac{\overline{\kappa}_{F}}{c^3}\frac{(f+\beta^2 )L-(1+f)\beta Q}{f-\beta ^2}\Biggr]\nonumber\\
        &&+\frac{\delta^2y}{4\pi r^2g_{00}\beta}(\Gamma \beta^2-\alpha^2_{\rm s})\frac{\overline{\kappa}_{E}}{c^3}\nonumber\\
        &&\times\Biggl[16\pi^2g_{00} r^2B-\frac{(1+\beta^2)Q-2\beta L}{f-\beta^2}\Biggr]\Biggr\},
\end{eqnarray}

\begin{eqnarray}
        \frac{dL}{dr}&=
        &\frac{\dot{M}}{4\pi r^2g_{00}c\beta}\Biggl\{\overline{\kappa}_{\rm E}\Biggl[16\pi^2g_{00}r^2B\nonumber\\
        &&-\frac{(1+\beta^2)Q-2\beta L}{f-\beta^2}\Biggr]\nonumber\\
        &&-\overline{\kappa}_{\rm F}\beta\frac{(f+\beta^2)L-(1+f)\beta Q}{f-\beta^2}\Biggr\}, \nonumber
\end{eqnarray}
\begin{equation} \label{0th-moment-re}
    ~
\end{equation}

\begin{eqnarray} 
    \frac{dQ}{dr}&=
    &\frac{1}{rg_{00}}\left(1-\frac{3r_{\rm S}}{2r}\right)\frac{(1-f)(1+\beta^2)Q-(1-f)2\beta L}{f-\beta^2}\nonumber\\
    &&+\frac{\dot{M}}{4\pi r^2g_{00}c\beta}\Biggl\{\beta\overline{\kappa}_{\rm E}\Biggl[16\pi^2g_{00}r^2B\nonumber\\
    &&-\frac{(1+\beta^2)Q-2\beta L}{f-\beta^2}\Biggr]\nonumber\\
    &&-\overline{\kappa}_{\rm F}\frac{(f+\beta^2)L-(1+f)\beta Q}{f-\beta^2}\Biggr\}, \nonumber 
\end{eqnarray}
\begin{equation} \label{1st-moment-re}
    ~
\end{equation}

\begin{equation} \label{tau-re}  
    \frac{d\tau}{dr}=-\overline{\kappa}_{\rm F}\frac{(1-\beta)}{4\pi r^2  g_{00} c\beta}\dot{M},
\end{equation}

\begin{equation} \label{bel-re}   
        \frac{\dot{M}c^2}{\delta ^2}y+L=\dot{E}. 
\end{equation}
Here, this Bernoulli equation is not a dependent one,
but use to determine the adiabatic sound speed, as follows.
\footnote{
Nonrelativistic limit of this Bernoulli equation is expressed
\begin{displaymath}
    \dot{M}c^2\left(1+\frac{1}{2}\beta^2+\frac{\alpha_{\rm s}^2}{\Gamma-1}-\frac{r_{\rm s}}{2r}\right)+L=\dot{E},
\end{displaymath}
when we plug $y\simeq 1-\frac{r_{\rm S}}{2r}+\frac{1}{2}\beta^2$, $\delta^{-2}\simeq 1+\frac{\alpha_{\rm s}^2}{\Gamma-1}$ in equation (\ref{bel-re}).
If we include the constant term $\dot{M}c^2$ on the right side, it has the same form as in Fukue (2014).
}

Next, we normalize these equations in terms of the nondimensional variables (\ref{non-d-variable}) and parameters (\ref{non-d-parameter}).
Then, equations (\ref{wind-beta-re}), (\ref{wind-alpha-re}), (\ref{0th-moment-re}), (\ref{1st-moment-re}), (\ref{tau-re}), and (\ref{bel-re})
are respectively normalized as follows:
\begin{eqnarray}\label{df1}
    \frac{d\beta}{d\hat{r}}&=
    &\frac{1}{y^2(\beta^2-\alpha ^2_{\rm s})}\Biggr\{\beta\Biggl[-\frac{1}{2\hat{r}^2}+\frac{2}{\hat{r}}\left(1-\frac{3}{4\hat{r}}\right)\alpha ^2_{\rm s}\nonumber\\
    &&+\frac{\delta^2y}{2\hat{r}^2g_{00}\gamma^2}\frac{(f+\beta^2 )\hat{L}-(1+f)\beta \hat{Q}}{f-\beta ^2}\Biggr]\nonumber\\
    &&+\frac{\delta^2y}{2\hat{r}^2g_{00}\gamma^2}(\Gamma-1)\epsilon\nonumber \\
    &&\times\left[\mathcal{B}g_{00} \hat{r}^2-\frac{(1+\beta^2)\hat{Q}-2\beta \hat{L}}{f-\beta^2}\right]\Biggr\},
\end{eqnarray}

\begin{eqnarray} \label{df2}
        \frac{d\alpha^2_{\rm s}}{d\hat{r}}=&
        &-\frac{(\Gamma-1)\delta^2}{y^2(\beta^2-\alpha ^2_{\rm s})}\Biggl\{\alpha ^2_{\rm s}\Biggl[-\frac{1}{2\hat{r}^2}+\frac{2}{\hat{r}}y^2\beta^2\nonumber\\
        &&+\frac{\delta^2y}{2\hat{r}^2g_{00}}\frac{(f+\beta^2 )\hat{L}-(1+f)\beta \hat{Q}}{f-\beta ^2}\Biggr]\nonumber\\
        &&+\frac{\delta^2y}{2\hat{r}^2g_{00}\beta}(\Gamma \beta^2-\alpha^2_{\rm s})\epsilon\nonumber\\
        &&\times\Biggl[\mathcal{B}g_{00} \hat{r}^2-\frac{(1+\beta^2)\hat{Q}-2\beta \hat{L}}{f-\beta^2}\Biggr]\Biggr\},
\end{eqnarray}

\begin{eqnarray} \label{df3}
    \frac{d\hat{L}}{d\hat{r}}&=
    &\frac{\dot{m}}{2\hat{r}^2g_{00}\beta}\Biggl\{\epsilon\left[\mathcal{B} g_{00}\hat{r}^2-\frac{(1+\beta^2)\hat{Q}-2\beta \hat{L}}{f-\beta^2}\right]\nonumber\\
    &&-\beta\frac{(f+\beta^2)\hat{L}-(1+f)\beta \hat{Q}}{f-\beta^2}\Biggr\},
\end{eqnarray}

\begin{eqnarray} \label{df4}
    \frac{d\hat{Q}}{d\hat{r}}&=
    &\frac{1}{\hat{r}g_{00}}\left(1-\frac{3}{2\hat{r}}\right)\frac{(1-f)(1+\beta^2)\hat{Q}-(1-f)2\beta \hat{L}}{f-\beta^2}\nonumber\\
    &&+\frac{\dot{m}}{2\hat{r}^2g_{00}\beta}\Biggl\{\beta\epsilon\left[\mathcal{B} g_{00}\hat{r}^2-\frac{(1+\beta^2)\hat{Q}-2\beta \hat{L}}{f-\beta^2}\right]\nonumber\\
    &&-\frac{(f+\beta^2)\hat{L}-(1+f)\beta \hat{Q}}{f-\beta^2}\Biggr\},
\end{eqnarray}

\begin{equation} \label{df5}
\frac{d\tau}{d\hat{r}}=-\frac{\dot{m}(1-\beta)}{2\hat{r}^2 g_{00}\beta},
\end{equation}

\begin{equation} \label{bel}
        \dot{e}=\frac{\dot{m}}{\delta ^2}y+\hat{L}. 
\end{equation}

Here, $\mathcal{B}$ and $\epsilon$ are the nondimensional blackbody intensity and photon destruction probability, respectively:
\begin{eqnarray}
    \mathcal{B} &=&\frac{16\pi^2r_{\rm S}^2 }{L_{\rm E}}B(T)=1.700\times10^{21}\frac{m}{\delta^8\Gamma^4}\alpha_{\rm s}^8\left(1+\frac{\kappa}{\sigma}\right),\\
    \frac{\kappa}{\sigma}&=&1.838\times10^{-27}\frac{\dot{m}}{m}\frac{\delta^7\Gamma^{3.5}}{\hat{r}^2 y\beta\alpha_{\rm s}^7},\\
    \epsilon &=&\frac{\overline{\kappa}_{\rm E}}{\overline{\kappa}_F}=\frac{\kappa}{\kappa+\sigma}=\frac{\frac{\kappa}{\sigma}}{1+\frac{\kappa}{\sigma}}. 
\end{eqnarray}

In addition, the adiabatic sound speed is expressed by using the nondimensional Bernoulli equation (\ref{bel}):
\begin{equation} \label{alpha}
    \alpha_{\rm s}^2=(\Gamma-1)\left(1-\frac{\dot{m}y}{\dot{e}-\hat{L}}\right).
\end{equation}
In order to determine the adiabatic sound speed,
we use this equation (\ref{alpha}), instead of (\ref{df2}),
for simplicity.

\section{Critical points}

The distance ($\hat{r}=\hat{r}_{\rm c}$), at which the flow speed ($\beta=\beta_{\rm c}$) is equal to the adiabatic sound speed ($\alpha_s=\alpha_{\rm s,c}$),
is called the {\it transonic point} ($\beta_{\rm c}=\alpha_{\rm s,c}$)
(the subscript c means `critical').
Wind equation (\ref{df1}) shows that the transonic points is also a {\it critical point} since at $\hat{r}=\hat{r}_{\rm c}$ the denominater of equation (\ref{df1}) vanishes.
Hence, in order for the transonic solution to exist,
the numerator of equation (\ref{df1}) must vanish at the critical point
simultaneously ({\it regularity condition}).

In this section, we first obtain and determine all the variables at $\hat{r}_{\rm c}$, calculate $d\beta/d\hat{r}|_{\rm c}$ 
by using the L'Hopital's rule in equation (\ref{df1}) at $\hat{r}=\hat{r}_{\rm c}$,
and examine the topology of the transonic/critical points.

Equation (\ref{df1}) is reexpressed as
\begin{equation} \label{N1-D}
    \frac{d\beta}{d\hat{r}}=\frac{\mathcal{N}_1}{\mathcal{D}},
\end{equation}
where
\begin{equation} \label{D}
    \mathcal{D} \equiv \beta ^2-\alpha _{\rm s}^2,
\end{equation}
\begin{eqnarray} \label{N1}    
        \mathcal{N} _1\nonumber&\equiv &f_1\nonumber\\
        &\equiv & \frac{1}{y^2}\Biggl\{\beta\Biggl[-\frac{1}{2\hat{r}^2}+\frac{2}{\hat{r}}\left(1-\frac{3}{4\hat{r}}\right)\alpha ^2_{\rm s}\nonumber\\
        &&+\frac{\delta^2y}{2\hat{r}^2g_{00}\gamma^2}\frac{(f+\beta ^2)\hat{L}-(1+f)\beta \hat{Q}}{f-\beta ^2}\Biggr]\nonumber\\
        &&+\frac{\delta^2y}{2\hat{r}^2g_{00}\gamma^2}(\Gamma-1)\epsilon\nonumber \\
        &&\times\left[\mathcal{B}g_{00} \hat{r}^2-\frac{(1+\beta^2)\hat{Q}-2\beta \hat{L}}{f-\beta^2}\right]\Biggr\}.
\end{eqnarray}
At the critical points, $\mathcal{D}$ and $\mathcal{N}_1$ must vanish simulteneously, as was stated.

Firstly, we derive a relation among the quantities at the critical point.
That is, imposing the condition of $\mathcal{D}|_{\rm c}=0$ ($\beta_{\rm c}=\alpha_{\rm s,c}$) on the Bernoulli equation (\ref{bel}), we can express $\beta_{\rm c}$ in terms of other quantities as
\begin{displaymath} 
    \beta_{\rm c}^6-(2\Gamma-1)\beta_{\rm c}^4+(\Gamma^2-1)\beta_{\rm c}^2-(\Gamma-1)^2\left[1-\frac{\dot{m}^2g_{00}}{(\dot{e}-\hat{L}_{\rm c})^2}\right]=0.
\end{displaymath}
\begin{equation} \label{cp-eq}
    ~
\end{equation} 
In order to simplify the expression, 
we introduce
\begin{equation}
    C\equiv 1-\frac{\dot{m}^2g_{00}}{(\dot{e}-\hat{L}_{\rm c})^2}
    = 1-\frac{\dot{m}^2}{(\dot{e}-\hat{L}_{\rm c})^2}
      \left( 1 - \frac{1}{\hat{r}_{\rm c}} \right),
\end{equation}
and solve equation (\ref{cp-eq}) to yield
\begin{eqnarray} 
        \beta_{\rm c}^2&=&\alpha_{\rm s,c}^2\nonumber\\
        &=&\frac{2\Gamma-1}{3}-\frac{1}{3}\Biggl\{-\frac{27C}{2}(\Gamma-1)^2+(1-2\Gamma)^3\nonumber\\
        &&-\frac{9(1-2\Gamma)(\Gamma^2-1)}{2}+\frac{1}{2}\Bigl[-4\left((1-2\Gamma)^2-3(\Gamma^2-1)\right)^3\nonumber\\
        &&+\Bigl(-27C(\Gamma-1)^2+2(1-2\Gamma)^3\nonumber\\
        &&-9(1-2\Gamma)(\Gamma^2-1)\Bigr)^2\Bigr]^{\frac{1}{2}}\Biggr\}^{\frac{1}{3}}\nonumber\\
        &&-\frac{(1-2\Gamma)^2-3(\Gamma^2-1)}{3}\Biggl\{-\frac{27C}{2}(\Gamma-1)^2+(1-2\Gamma)^3\nonumber\\
        &&-\frac{9(1-2\Gamma)(\Gamma^2-1)}{2}+\frac{1}{2}\Bigl[-4\left((1-2\Gamma)^2-3(\Gamma^2-1)\right)^3\nonumber\\
        &&+\Bigl(-27C(\Gamma-1)^2+2(1-2\Gamma)^3\nonumber\\
        &&-9(1-2\Gamma)(\Gamma^2-1)\Bigr)^2\Bigr]\Biggr\}^{-\frac{1}{3}}.\nonumber      
\end{eqnarray}
\begin{equation} \label{beta-c-gamma}
    ~
\end{equation}
For example, when $\Gamma=4/3$, equation (\ref{beta-c-gamma}) becomes 
\begin{eqnarray}
        \beta_{\rm c}^2&=&\alpha_{\rm s, c}^2\nonumber\\
        &=&\frac{5}{9}-\frac{1}{3}\left[-\frac{3C}{2}+\frac{1}{2}\sqrt{\left(\frac{65}{27}-3C\right)^2-\frac{256}{729}}+\frac{65}{54}\right]^{\frac{1}{3}}\nonumber\\
        &&-\frac{4}{27}\left[-\frac{3C}{2}+\frac{1}{2}\sqrt{\left(\frac{65}{27}-3C\right)^2-\frac{256}{729}}+\frac{65}{54}\right]^{-\frac{1}{3}},\nonumber
\end{eqnarray}
\begin{equation}\label{beta-c-sol4/3}
    ~
\end{equation}
and when $\Gamma=5/3$, it is
\begin{eqnarray} 
    \beta_{\rm c}^2&=&\alpha_{\rm s, c}^2\nonumber\\
    &=&\frac{7}{9}-\frac{1}{3}\left[-6C
    +\frac{1}{2}\sqrt{\left(\frac{322}{27}-12C\right)^2-\frac{4}{729}}+\frac{161}{27}\right]^{\frac{1}{3}}\nonumber \\
    &&-\frac{1}{27}\left[-6C
    +\frac{1}{2}\sqrt{\left(\frac{322}{27}-12C\right)^2-\frac{4}{729}}+\frac{161}{27}\right]^{-\frac{1}{3}}. \nonumber 
\end{eqnarray}
\begin{equation}\label{beta-c-sol5/3}
    ~
\end{equation}

It should be stressed that equations (\ref{beta-c-gamma}), (\ref{beta-c-sol4/3}) {and (\ref{beta-c-sol5/3})} can have physical solutions only if $0\leq C\leq 1$.
If $C=0$, $\beta_{\rm c}=0$, and if $C=1$, $\beta_{\rm c}=\sqrt{\Gamma-1}$.  
In particular, at the horizon,
where $g_{00}=0$, $C=1$, therefore $\beta_{\rm c}=\sqrt{\Gamma-1}$. 
Another condition for the quantities at the critical point
is obtained from the regularity condition: $\mathcal{N}_1|_{\rm c}=0$.
Other relations for, e.g., radiation pressure $\hat{Q}_{\rm c}$ and optical depth $\tau_{\rm c}$ can be obtained from the remained equations as below. 

Next, applying the L'Hopital's rule in equation (\ref{N1-D}) at $\hat{r}=\hat{r}_{\rm c}$, 
$\left.d\beta/d\hat{r}\right|_{\rm c}$ is written as
\begin{eqnarray} 
    &&\left.\frac{d\beta}{d\hat{r}}\right|_{\rm c} \nonumber\\
    &&=\frac{\left.\frac{\partial \mathcal{N}_1}{\partial \hat{r}}\right|_{\rm c}
    +\left.\frac{\partial \mathcal{N}_1}{\partial \beta }\right|_{\rm c}\left.\frac{d\beta}{d\hat{r}}\right|_{\rm c}
    +\left.\frac{\partial \mathcal{N}_1}{\partial \hat{L} }\right|_{\rm c}\left.\frac{d\hat{L}}{d\hat{r}}\right|_{\rm c}
    +\left.\frac{\partial \mathcal{N}_1}{\partial \hat{Q}}\right|_{\rm c}\left.\frac{d\hat{Q}}{d\hat{r}}\right|_{\rm c}
    +\left.\frac{\partial \mathcal{N}_1}{\partial \tau }\right|_{\rm c}\left.\frac{d\tau}{d\hat{r}}\right|_{\rm c}}
    {\left.\frac{\partial \mathcal{D}}{\partial \hat{r}}\right|_{\rm c}
    +\left.\frac{\partial \mathcal{D} }{\partial \beta }\right|_{\rm c}\left.\frac{d\beta}{d\hat{r}}\right|_{\rm c}
    +\left.\frac{\partial \mathcal{D} }{\partial \hat{L} }\right|_{\rm c}\left.\frac{d\hat{L}}{d\hat{r}}\right|_{\rm c}
    +\left.\frac{\partial \mathcal{D} }{\partial \hat{Q}}\right|_{\rm c}\left.\frac{d\hat{Q}}{d\hat{r}}\right|_{\rm c}
    +\left.\frac{\partial \mathcal{D} }{\partial \tau }\right|_{\rm c}\left.\frac{d\tau}{d\hat{r}}\right|_{\rm c}}.\nonumber
\end{eqnarray}
\begin{equation} \label{dbeta-c}
    ~
\end{equation}

In order to calculate equation (\ref{dbeta-c}), 
equations (\ref{df2}), (\ref{df3}), (\ref{df4}), and (\ref{df5}) are also tranformed in the same way: 
\begin{equation} \label{N2-D}
        \frac{d\alpha^2_{\rm s}}{d\hat{r}}=\frac{\mathcal{N}_2}{\mathcal{D}}
\end{equation}
\begin{equation}
        \frac{d\hat{L}}{d\hat{r}}=\frac{\mathcal{N}_3}{\mathcal{D}},
\end{equation}
\begin{equation}    
    \frac{d\hat{Q}}{d\hat{r}}=\frac{\mathcal{N}_4}{\mathcal{D}},
\end{equation}
\begin{equation}  
    \frac{d\tau}{d\hat{r}}=\frac{\mathcal{N}_5}{\mathcal{D}},
\end{equation}
where 
\begin{eqnarray} \label{N2}
    \mathcal{N} _2&\equiv&f_2\nonumber\\
    &\equiv &-(\Gamma-1)\frac{\delta^2}{y^2}\Biggl\{\alpha ^2_{\rm s}\Biggl[-\frac{1}{2\hat{r}^2}+\frac{2}{\hat{r}}y^2\beta^2\nonumber\\
    &&+\frac{\delta^2y}{2\hat{r}^2g_{00}}\frac{(f+\beta )\hat{L}-(1+f)\beta \hat{Q}}{f-\beta ^2}\Biggr]\nonumber\\
    &&+\frac{\delta^2y}{2\hat{r}^2g_{00}}\frac{1}{\beta}(\Gamma \beta^2-\alpha^2_{\rm s})\epsilon \nonumber \\
    &&\times \Biggl[\mathcal{B}g_{00} \hat{r}^2-\frac{(1+\beta^2)\hat{Q}-2\beta \hat{L}}{f-\beta^2}\Biggr]\Biggr\},
\end{eqnarray}
\begin{eqnarray}
    \mathcal{N} _3&\equiv&\mathcal{D}f_3\nonumber\\
    &\equiv &\frac{(\beta ^2-\alpha _{\rm s}^2)\dot{m}}{2\hat{r}^2g_{00}\beta}\Biggl\{\epsilon\Biggl[\mathcal{B} g_{00}\hat{r}^2-\frac{(1+\beta^2)\hat{Q}-2\beta \hat{L}}{f-\beta^2}\Biggr]\nonumber\\
    &&-\beta\frac{(f+\beta^2)\hat{L}-(1+f)\beta \hat{Q}}{f-\beta^2}\Biggr\},
\end{eqnarray}
\begin{eqnarray}
    \mathcal{N} _4&\equiv &\mathcal{D}f_4\nonumber\\
    &\equiv &(\beta ^2-\alpha _{\rm s}^2)\Biggl\{\frac{1}{\hat{r}g_{00}}\left(1-\frac{3}{2\hat{r}}\right)\nonumber\\
    &&\times\frac{(1-f)(1+\beta^2)\hat{Q}-(1-f)2\beta \hat{L}}{f-\beta^2}\nonumber\\
    &&+\frac{\dot{m}}{2\hat{r}^2g_{00}\beta}\Biggl[\epsilon\left(\mathcal{B} g_{00}\hat{r}^2-\frac{(1+\beta^2)\hat{Q}-2\beta \hat{L}}{f-\beta^2}\right)\nonumber\\
    &&-\frac{(f+\beta^2)\hat{L}-(1+f)\beta \hat{Q}}{f-\beta^2}\Biggr]\Biggr\},
\end{eqnarray}
\begin{eqnarray}  
    \mathcal{N} _5&\equiv &\mathcal{D}f_5\nonumber\\
    &\equiv &-(\beta ^2-\alpha _{\rm s}^2)\frac{\dot{m}(1-\beta)}{2\hat{r}^2 g_{00}\beta},
\end{eqnarray}
although in this paper we do not use equation (\ref{N2-D}),
but use the Bernoulli equation (\ref{bel}).

Now, we define the eigenvalue matrix $\Lambda$ as follows:
\begin{eqnarray}\label{matrix}
    \Lambda  &\equiv&
     \left(
    \begin{array}{ccccc}
      \lambda _{11} &\lambda _{12} &\lambda _{13} &\lambda _{14} &\lambda _{15} \\ 
      \lambda _{21} &\lambda _{22} &\lambda _{23} &\lambda _{24} &\lambda _{25} \\
      \lambda _{31} &\lambda _{32} &\lambda _{33} &\lambda _{34} &\lambda _{35} \\
      \lambda _{41} &\lambda _{42} &\lambda _{43} &\lambda _{44} &\lambda _{45} \\
      \lambda _{51} &\lambda _{52} &\lambda _{53} &\lambda _{54} &\lambda _{55}
    \end{array}
    \right)\nonumber\\
    &\equiv& \left(
      \begin{array}{ccccc}
        \left.\frac{\partial \mathcal{D}}{\partial \hat{r}}\right|_{\rm c} &  \left.\frac{\partial \mathcal{D}}{\partial \beta}\right|_{\rm c} &  \left.\frac{\partial \mathcal{D}}{\partial \hat{L}}\right|_{\rm c} &  \left.\frac{\partial \mathcal{D}}{\partial \hat{Q}}\right|_{\rm c} &  \left.\frac{\partial \mathcal{D}}{\partial \tau}\right|_{\rm c} \\
        \left.\frac{\partial \mathcal{N}_1 }{\partial \hat{r}}\right|_{\rm c} & \left.\frac{\partial \mathcal{N}_1 }{\partial \beta}\right|_{\rm c} & \left.\frac{\partial \mathcal{N}_1 }{\partial \hat{L}}\right|_{\rm c} & \left.\frac{\partial \mathcal{N}_1 }{\partial \hat{Q}}\right|_{\rm c} & \left.\frac{\partial \mathcal{N}_1 }{\partial \tau}\right|_{\rm c} \\
        \left.\frac{\partial \mathcal{N}_3 }{\partial \hat{r}}\right|_{\rm c} & \left.\frac{\partial \mathcal{N}_3 }{\partial \beta}\right|_{\rm c} & \left.\frac{\partial \mathcal{N}_3 }{\partial \hat{L}}\right|_{\rm c} & \left.\frac{\partial \mathcal{N}_3 }{\partial \hat{Q}}\right|_{\rm c} & \left.\frac{\partial \mathcal{N}_3 }{\partial \tau}\right|_{\rm c} \\
        \left.\frac{\partial \mathcal{N}_4 }{\partial \hat{r}}\right|_{\rm c} & \left.\frac{\partial \mathcal{N}_4 }{\partial \beta}\right|_{\rm c} & \left.\frac{\partial \mathcal{N}_4 }{\partial \hat{L}}\right|_{\rm c} & \left.\frac{\partial \mathcal{N}_4 }{\partial \hat{Q}}\right|_{\rm c} & \left.\frac{\partial \mathcal{N}_4 }{\partial \tau}\right|_{\rm c} \\
        \left.\frac{\partial \mathcal{N}_5 }{\partial \hat{r}}\right|_{\rm c} & \left.\frac{\partial \mathcal{N}_5 }{\partial \beta}\right|_{\rm c} & \left.\frac{\partial \mathcal{N}_5 }{\partial \hat{L}}\right|_{\rm c} & \left.\frac{\partial \mathcal{N}_5 }{\partial \hat{Q}}\right|_{\rm c} & \left.\frac{\partial \mathcal{N}_5 }{\partial \tau}\right|_{\rm c}          \end{array}
    \right).\nonumber\\
    &&
\end{eqnarray}
After simple but lengthy calculations,
this matrix is found to be expressed as
\begin{eqnarray}
    \Lambda 
    &=&\left(
      \begin{array}{ccccc}
        \left.\frac{\partial \mathcal{D}}{\partial \hat{r}}\right|_{\rm c}&  \left.\frac{\partial \mathcal{D}}{\partial \beta}\right|_{\rm c} & \left.\frac{\partial \mathcal{D}}{\partial \hat{L}}\right|_{\rm c}& 0 &  0 \\
        \left.\frac{\partial f_1 }{\partial \hat{r}}\right|_{\rm c} & \left.\frac{\partial f_1 }{\partial \beta}\right|_{\rm c} & \left.\frac{\partial f_1 }{\partial \hat{L}}\right|_{\rm c} & \left.\frac{\partial f_1 }{\partial \hat{Q}}\right|_{\rm c} & \left.\frac{\partial f_1 }{\partial \tau}\right|_{\rm c} \\
        \left.\frac{\partial \mathcal{D}}{\partial \hat{r}}\right|_{\rm c}\left.f_3\right|_{\rm c} & \left.\frac{\partial \mathcal{D}}{\partial \beta}\right|_{\rm c}\left.f_3\right|_{\rm c} & \left.\frac{\partial \mathcal{D}}{\partial \hat{L}}\right|_{\rm c}\left.f_3\right|_{\rm c} & 0 & 0 \\
        \left.\frac{\partial \mathcal{D}}{\partial \hat{r}}\right|_{\rm c}\left.f_4\right|_{\rm c} & \left.\frac{\partial \mathcal{D}}{\partial \beta}\right|_{\rm c}\left.f_4\right|_{\rm c} & \left.\frac{\partial \mathcal{D}}{\partial \hat{L}}\right|_{\rm c}\left.f_4\right|_{\rm c} & 0 & 0 \\
        \left.\frac{\partial \mathcal{D}}{\partial \hat{r}}\right|_{\rm c}\left.f_5\right|_{\rm c} & \left.\frac{\partial \mathcal{D}}{\partial \beta}\right|_{\rm c}\left.f_5\right|_{\rm c} & \left.\frac{\partial \mathcal{D}}{\partial \hat{L}}\right|_{\rm c}\left.f_5\right|_{\rm c} & 0 & 0 
      \end{array}
    \right)\nonumber\\
    &=&\left(
    \begin{array}{ccccc}
      \lambda _{11} &\lambda _{12} &\lambda _{13} & 0 & 0\\
      \lambda _{21} &\lambda _{22} &\lambda _{23} &\lambda _{24} &\lambda _{25} \\
      \lambda _{11}\left.f_3\right|_{\rm c} &\lambda _{12}\left.f_3\right|_{\rm c} & \lambda _{13}\left.f_3\right|_{\rm c} & 0 & 0\\
      \lambda _{11}\left.f_4\right|_{\rm c} &\lambda _{12}\left.f_4\right|_{\rm c} & \lambda _{13}\left.f_4\right|_{\rm c} & 0 & 0\\
      \lambda _{11}\left.f_5\right|_{\rm c} &\lambda _{12}\left.f_5\right|_{\rm c} & \lambda _{13}\left.f_5\right|_{\rm c} & 0 & 0
    \end{array}
    \right).\nonumber
\end{eqnarray}
\begin{equation}\label{matrix2} 
    ~
\end{equation}
Thus, fortunately,
the eigenvalue equation reduces to the quadratic one, as shown below.

Using the components of matrix $\Lambda$, equation (\ref{dbeta-c}) is further rewritten as
\begin{equation}
    \left.\frac{d\beta}{d\hat{r}}\right|_{\rm c}
    =\frac{\lambda_{21}+\lambda_{22}\left.\frac{d\beta}{d\hat{r}}\right|_{\rm c}
    +\lambda_{23}\left.f_3\right|_{\rm c}+\lambda_{24}\left.f_4\right|_{\rm c}
    +\lambda_{25}\left.f_5\right|_{\rm c}}
    {\lambda_{11}+\lambda_{12}\left.\frac{d\beta}{d\hat{r}}\right|_{\rm c}
    +\lambda_{13}\left.f_3\right|_{\rm c}},
\end{equation}
and finally solved as
\begin{eqnarray} \label{dbeta-dr-c}
    \left.\frac{d\beta}{d\hat{r}}\right|_{\rm c}&=&
    \frac{1}{2\lambda_{12}}\Biggl[-\left(\lambda_{11}-\lambda_{22}+\lambda_{13}\left.f_3\right|_{\rm c}\right)\nonumber\\
    &&\pm\sqrt{\left(\lambda_{11}-\lambda_{22}+\lambda_{13}\left.f_3\right|_{\rm c}\right)^2+4\lambda_{12}\xi}\Biggr],
\end{eqnarray}
where
\begin{equation}
    \xi \equiv \lambda_{21}+\lambda_{23}\left.f_3\right|_{\rm c}+\lambda_{24}\left.f_4\right|_{\rm c}+\lambda_{25}\left.f_5\right|_{\rm c}.
\end{equation}
When equation (\ref{dbeta-dr-c}) has a real solution, it gives a linear approximated solution in the vicinity of the critical point.   

Thirdly, we determine the types of critical points by the eigenvalue of the matrix $\Lambda$.
The eigenvalue equation of $\Lambda$ is now
\begin{displaymath} 
    \lambda^2-(\lambda_{11}+\lambda_{22}+\lambda_{13}\left.f_{3}\right|_{\rm c})\lambda
    +(\lambda_{11}+\lambda_{13}\left.f_{3}\right|_{\rm c})\lambda_{22}-\xi\lambda_{12}=0.
\end{displaymath}
\begin{equation} \label{eigenvalue-eq}
    ~
\end{equation}

Depending on what solution this equation has, we can determine the types of critical points.
If equation (\ref{eigenvalue-eq}) has two real solutions with different signs, the critical points are of the {\it saddle type}, 
two real solutions with the same sign of the {\it nodal one}, 
and two complex solutions of the {\it spiral one}.
Since the nodal type is always a deceleration solution and the spiral type is not a physical solution in the present case,
the most suitable transonic solution is that passes the critical point of the saddle type.

\begin{figure}
    \includegraphics[width=80mm,clip]{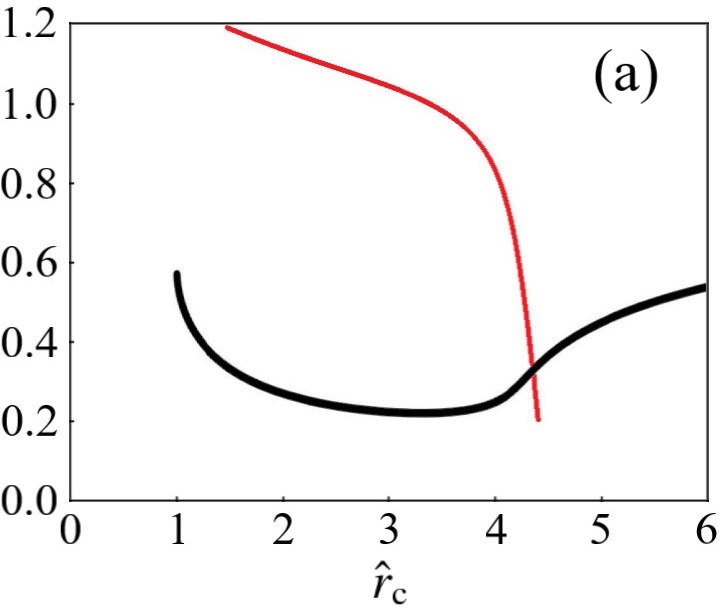}
    \includegraphics[width=80mm,clip]{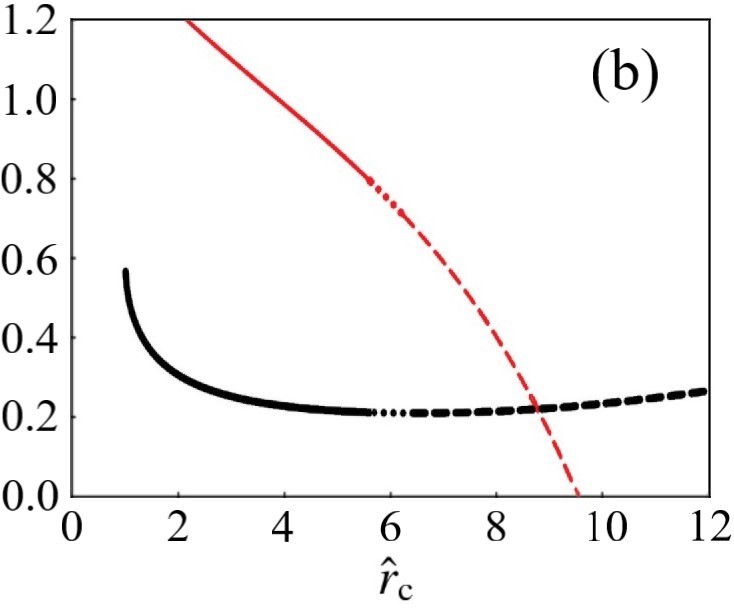}
    \includegraphics[width=80mm,clip]{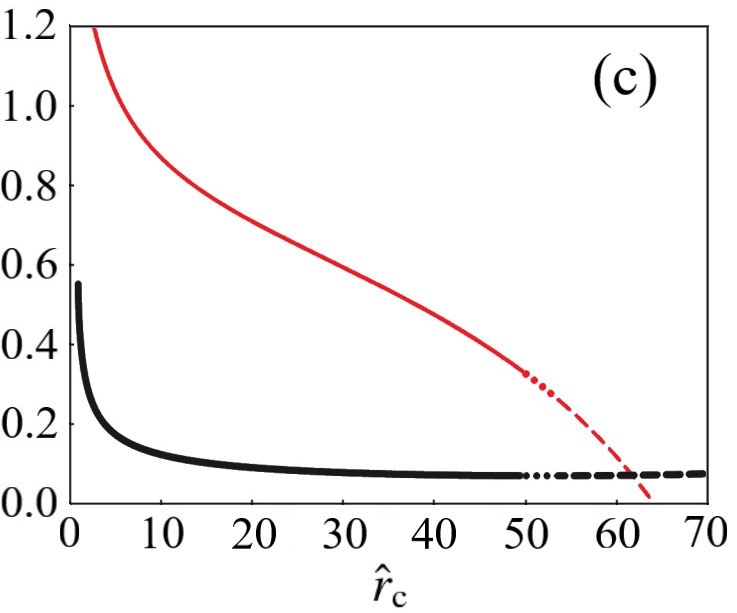}
    \caption{
    Critical curves between $\hat{r}_{\rm c}$, $\beta_{\rm c}$, $\hat{L}_{\rm c}$;
    (a) $\dot{m}=2$, $\dot{e}=3$, $m=10^8$, $\Gamma=4/3$, $\hat{Q}_{\rm c}=1$, $\tau_{\rm c}=1$,
    (b) $\dot{m}=10$, $\dot{e}=11.5$, $m=10^8$, $\Gamma=4/3$, $\hat{Q}_{\rm c}=1$, $\tau_{\rm c}=5$,
    (c) $\dot{m}=100$, $\dot{e}=101$, $m=10^8$, $\Gamma=4/3$, $\hat{Q}_{\rm c}=1$, $\tau_{\rm c}=5$.
    Black thick curves denote $\beta_{\rm c}$, while red thin ones $L_{\rm c}$.
    Solid curves mean saddle type, dotted ones nodal type, and dashed ones spiral type. 
    }
\end{figure} 

\begin{figure}
    \includegraphics[width=81.2mm,clip]{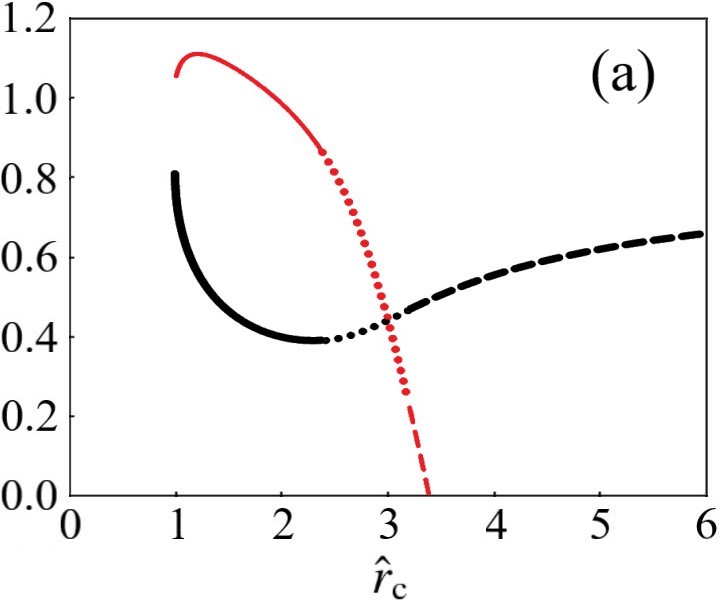}
    \includegraphics[width=81.2mm,clip]{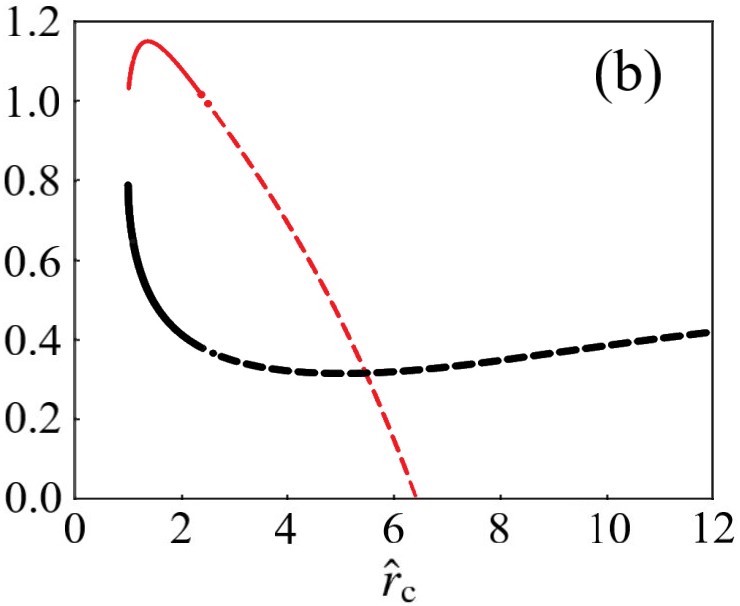}
    \caption{
        Critical curves between $\hat{r}_{\rm c}$, $\beta_{\rm c}$, and $\hat{L}_{\rm c}$;
        (a) $\dot{m}=2$, $\dot{e}=3$, $m=10^8$, $\Gamma=5/3$, $\hat{Q}_{\rm c}=1$, $\tau_{\rm c}=1$, 
        (b) $\dot{m}=10$, $\dot{e}=11.5$, $m=10^8$, $\Gamma=5/3$, $\hat{Q}_{\rm c}=1$, $\tau_{\rm c}=5$.
        Black thick curves denote $\beta_{\rm c}$, while red thin ones $\hat{L}_{\rm c}$.
        Solid curves mean saddle type, dotted ones nodal type, and dashed ones spiral type. }
\end{figure}

In Fig. 1, the loci of critical points and their types are shown
for several typical parameters;
Fig. 1a for 
$\dot{m}=2$, $\dot{e}=3$, $m=10^8$, $\Gamma=4/3$, $\hat{Q}_{\rm c}=1$, $\tau_{\rm c}=1$,
Fig. 1b for 
$\dot{m}=10$, $\dot{e}=11.5$, $m=10^8$, $\Gamma=4/3$, $\hat{Q}_{\rm c}=1$, $\tau_{\rm c}=5$,
and Fig. 1c for
$\dot{m}=100$, $\dot{e}=101$, $m=10^8$, $\Gamma=4/3$, $\hat{Q}_{\rm c}=1$, $\tau_{\rm c}=5$.
Black thick curves denote $\beta_{\rm c}$, while red thin ones $L_{\rm c}$.
Solid curves mean saddle type, dotted ones nodal type, and dashed ones spiral type. 

When the mass-loss rate $\dot{m}$ is large (Fig. 1c),
the flow velocity at critical points becomes generally low,
since the loaded mass is large.
In such a case, the loci and types of critical points
resemble those obtained in the nonrelativistic regime (Fukue 2014).
That is,
the saddle type appears in the inner branch,
while the types in the outer region are nodal or spiral.
Especially, if we drop the radiation drag term,
and give the same mass-loss rates as those of Fukue (2014),
say $\dot{m}=10^5$ for nova winds or $\dot{m}=10^3$ for neutron star winds,
the critical curves almost coincide to those of Fukue (2014)
in the nonrelativistic regime.
However, 
at far from the center,
the velocity at the critical points diverges in the nonrelativistic case,
while in the general relativistic case it approaches the relativistic limit 
of the adiabatic sound speed,
$\beta_{\rm c} = \alpha_{\rm s, c} \rightarrow 1/\sqrt{3}$.

When the mass-loss rate becomes small (Figs. 1a and 1b),
the flow velocity at critical points becomes high.
In such cases, the general behavior is similar to the nonrelativistic regime,
but the loci and types are rather different.
That is,
the loci of critical points move inward,
and further nodal and spiral types disappear.
Anyway,
the loci and types of critical points,
and the velocity and luminosity there depend on the 
various parameters (see Fig. 3 below).

Before it,
we briefly mention the case of $\Gamma=5/3$.
As was shown in, e.g., Holzer and Axford (1970),
in the nonrelativistic regime
the critical points do not exist for $\Gamma=5/3$.
Hence, in Fukue (2014) and in the present study
we set $\Gamma=4/3$.
However, in contrast to the nonrelativistic regime,
in the present relativistic regime
the critical points do exist even for $\Gamma=5/3$,
as is shown in Fig. 2.
Fig. 2 shows the critical curves for the same parameters as in Figs. 1a and 1b, but $\Gamma$ is $5/3$.
As is seen in Fig. 2,
the loci and types of critical points are somewhat similar
to those in Fig. 1, although they are quantitatively different.
The existence of critical points for $\Gamma=5/3$ is 
due to the relativistic effect,
especially the relativistic gravity.
In the nonrelativistic case (Holzer \& Axford 1970),
the case of $\Gamma=5/3$ is marginal,
and the critical points exist for $\Gamma<5/3$.
Hence,
if the Newtonian gravity may slightly change,
the critical points can exist.

\begin{figure}
    \includegraphics[width=83mm,clip]{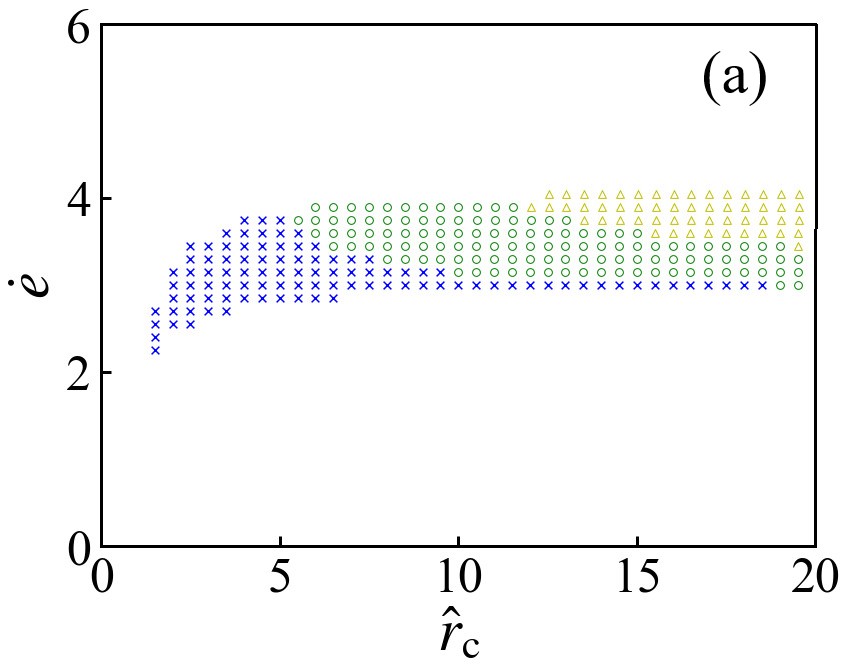}
    \includegraphics[width=83mm,clip]{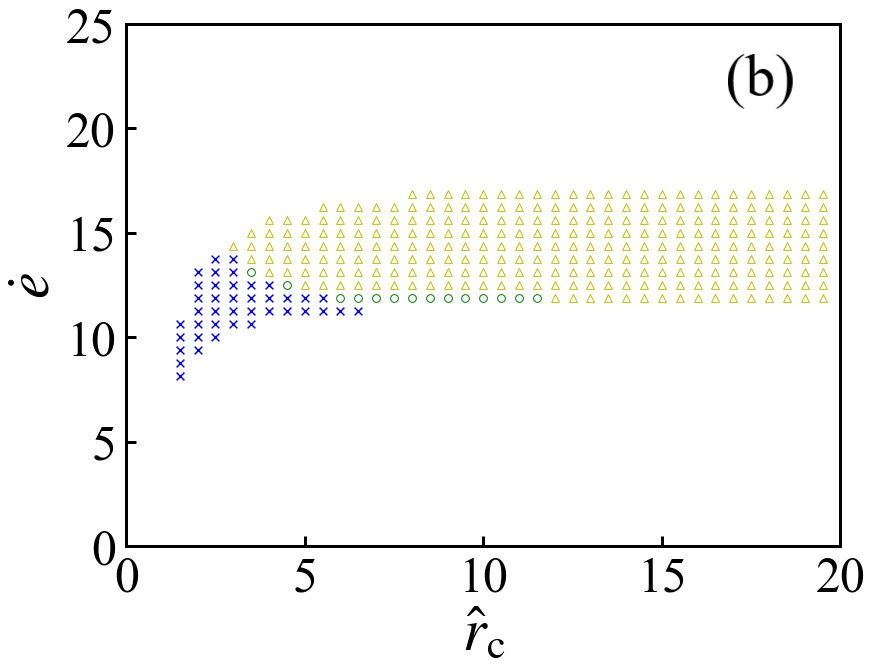}
    \includegraphics[width=83mm,clip]{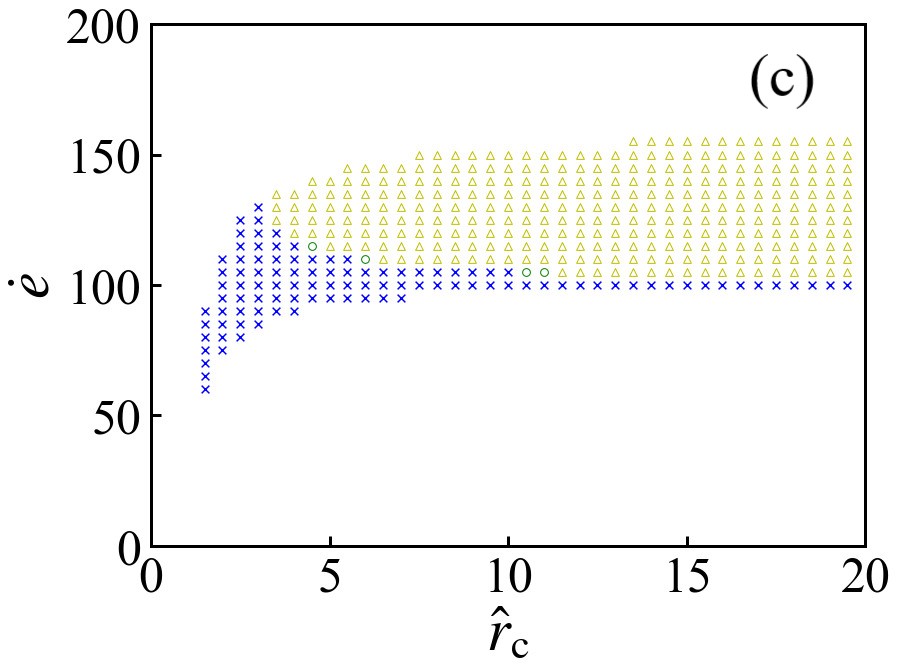}
    \caption{
Types of critical points
in the $\hat{r}_{\rm c}$-$\dot{e}$ parameter space.
Blue crosses mean saddle type, 
green open circles nodal type, 
and yellow open triangles spiral type.
No critical points appear in the margin no mark region.
Parameters are
(a) 
$\dot{m}=2$, $\hat{L}_{\rm c}=1$, $\tau_{\rm c}=1$, $m=10^8$, $\Gamma=4/3$,
(b)
$\dot{m}=10$, $\hat{L}_{\rm c}=2$, $\tau_{\rm c}=10$, $m=10^8$, $\Gamma=4/3$,
(c)
$\dot{m}=100$, $\hat{L}_{\rm c}=2$, $\tau_{\rm c}=20$, $m=10^8$, $\Gamma=4/3$.
    }
\end{figure}

It should be mentioned the limitation of constant $\Gamma$ assumption.
In the vicinity of the center, 
the ratio of specific heats should approach $4/3$ due to the high temperature of the gas, 
and the velocity at the critical point should also approach the relativistic limit of the sound speed ($\alpha_{\rm s}=1/\sqrt{3}$).
As is seen in Fig. 2, however, the velocity of the critical point near to the center (and far from the center) exceeds this limit of $1/\sqrt{3}$,
since we fixed $\Gamma=5/3$ in this calculation.
Hence, in order to treat and solve this problem,
we must use the variable $\Gamma(T)$,
which depends on the gas temperature (e.g., Kato et al. 2008).

We now summarize the types of critical points
in the $\hat{r}_{\rm c}$-$\dot{e}$ parameter space.
In Fig. 3,
blue crosses mean saddle type, 
green open circles nodal type, 
and yellow open triangles spiral type.
No critical points appear in the margin no mark region.
Parameters are
(a) 
$\dot{m}=2$, $\hat{L}_{\rm c}=1$, $\tau_{\rm c}=1$, $m=10^8$, $\Gamma=4/3$,
(b)
$\dot{m}=10$, $\hat{L}_{\rm c}=2$, $\tau_{\rm c}=10$, $m=10^8$, $\Gamma=4/3$,
(c)
$\dot{m}=100$, $\hat{L}_{\rm c}=2$, $\tau_{\rm c}=20$, $m=10^8$, $\Gamma=4/3$.
 
As is seen in Fig. 3,
types of critical points roughly align
from the inner region (saddle),
via the middle one (nodal),
to the outer one (spiral).
The divided pattern is similar,
although $\dot{e}$ becomes large as $\dot{m}$ becomes large.
For example, the saddle type critical points appear 
relatively close to the central black hole.
As a result,
the transonic flows are accelerated
in the vicinity of the central black hole,
as shown in the next section.

In addition,
when the value of $\dot{e}$ is large, the flow is always supersonic,
while it is always subsonic when the value of $\dot{e}$ is small.
This is the reason that
no critical points appear in high and low $\dot{e}$ regions.

Finally, we should note that 
the existence of the saddle-type critical point 
does not always mean that a physically-reasonable solution can be obtained.
For example,
if the radiation pressure and radiation drag is too large at large distance,
the flow may be decelerated to the subsonic speed.
Hence,
it is important to choose the appropriate parameters
at the critical point.

\section{Transonic solutions}

For appropriate parameters,
from the critical point $\hat{r}_{\rm c}$,
we integrate equations (\ref{df1}), (\ref{df3}), (\ref{df4}), (\ref{df5}), and (\ref{alpha}) instead of (\ref{df2}),
inward and outward  using the 4-th Runge-Kutta method
to solve and obtain transonic solutions.
In this section
we show several examples of transonic solutions.

As appropriate parameters,
we restrict several parameters, as mentioned in section 2,
bearing in mind UFOs, especially PG1211+143 (Tombesi et al. 2012).
Namely, the mass of the central supermassive black hole
is assumed to be $m=10^8$.
The nondimensional mass-loss rate 
is set to be $\dot{m} \sim 1$--100.
Furthermore,
the typical outflow velocity of UFOs is $\beta_{\infty}\sim 0.1$--0.3, 
and the typical luminosity is assumed to be on the order of Eddington luminosity; $\hat{L}_{\infty}\sim 1$.
Under these restricted parameters, 
the Bernoulli equation (\ref{bel-re}) is estimated at infinity as
\begin{eqnarray} \label{dote-infty}
    \dot{e} &\sim& \dot{m}\gamma_{\infty}+\hat{L}_{\infty}
\nonumber \\
    &\sim& \dot{m}\left( 1+\frac{1}{2}\beta^2_{\infty} \right)+\hat{L}_{\infty}
\nonumber \\
    &\sim& (1.005{\rm -}1.05)\dot{m}+1.
\end{eqnarray}
This relation restricts the range of parameter $\dot{e}$.
For example, when $\dot{m}=2$, $\dot{e}\sim 3.01$--3.1.

\begin{figure}
   \includegraphics[width=80mm,clip]{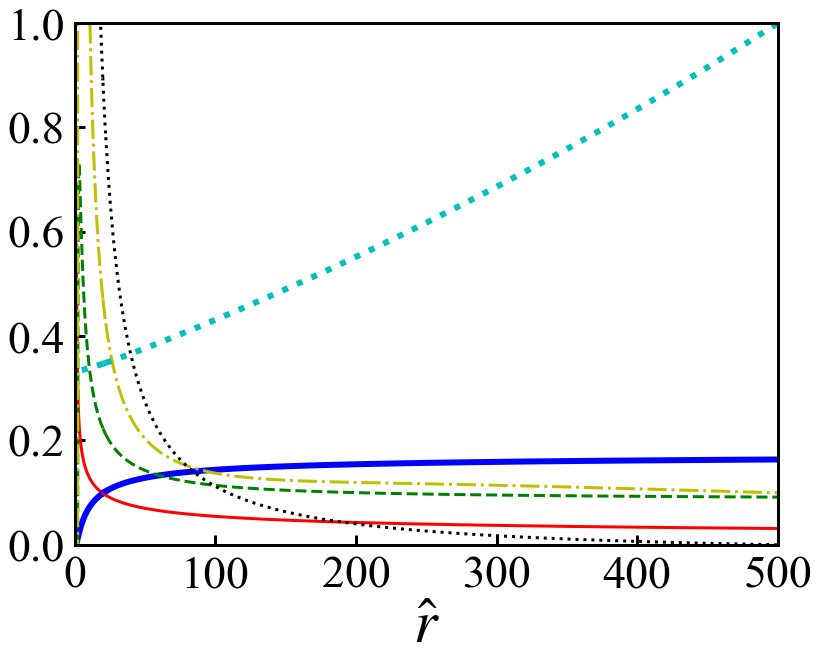}
   \caption{
    Typical transonic solutions.
Parameters are
$\dot{m}=100$, $\dot{e}=102$, $m=10^8$, $\Gamma=4/3$, $\hat{r}_{\rm c}=20$, $\hat{L}_{\rm c}=1.118$, $Q_{\rm c}=2.363$, $\tau_{\rm c}=17.9$.
A blue thick solid curve means $\beta$, a red solid one $\alpha_{\rm s}$, a green dashed one $\hat{L}/5$,
a yellow chain-dotted one $\hat{Q}/5$, a black dotted one $\tau/20$, and a cyan thick dotted one $f(\beta,\tau)$.
    }
\end{figure}

Fig. 4 shows a typical example of transonic solutions.
Parameters are
$\dot{m}=100$, $\dot{e}=102$, $m=10^8$, $\Gamma=4/3$, $\hat{r}_{\rm c}=20$, $\hat{L}_{\rm c}=1.118$, $Q_{\rm c}=2.363$, $\tau_{\rm c}=17.9$.
A blue thick solid curve means $\beta$, a red solid one $\alpha_{\rm s}$, a green dashed one $\hat{L}/5$,
a yellow chain-dotted one $\hat{Q}/5$, a black dotted one $\tau/20$, and a cyan thick dotted one $f(\beta,\tau)$.

As is seen in Fig. 4,
the wind is mainly accelerated at around the critical point,
and approaches a terminal speed,
while the luminosity becomes almost constant of $\hat{L}\sim 0.457$.
In this case
the wind terminal speed is about $0.163~c$,
which is preferable for UFOs, as assumed.
It should be noted that in this case
the optical depth vanishes at around $r \sim 500~r_{\rm S}$,
which is the wind top,
and the Eddington factor approaches unity there.

It should be noted that
the flow becomes optically thin at the large distance $r$
since the gas density decreases.
Although the variable Eddington factor can be applied 
in the optically thin regime,
the diffusion approximation itself becomes inappropriate there.
Hence, the present transonic solution is not appropriate at large $r$.

In addition,
in the realistic black hole wind, 
such as UFOs,
there may exist a luminous accretion disc surrounding a black hole.
As a result, except for the inner optically thick flow,
the outer optically thin part would be affected
by the disc radiation.
Hence, the spherically symmetric present model
would be modified in the region at large $r$ again.

Parameter dependence of transonic solutions
is depicted in Figs. 5 and 6.

\begin{figure}
    \includegraphics[width=80mm,clip]{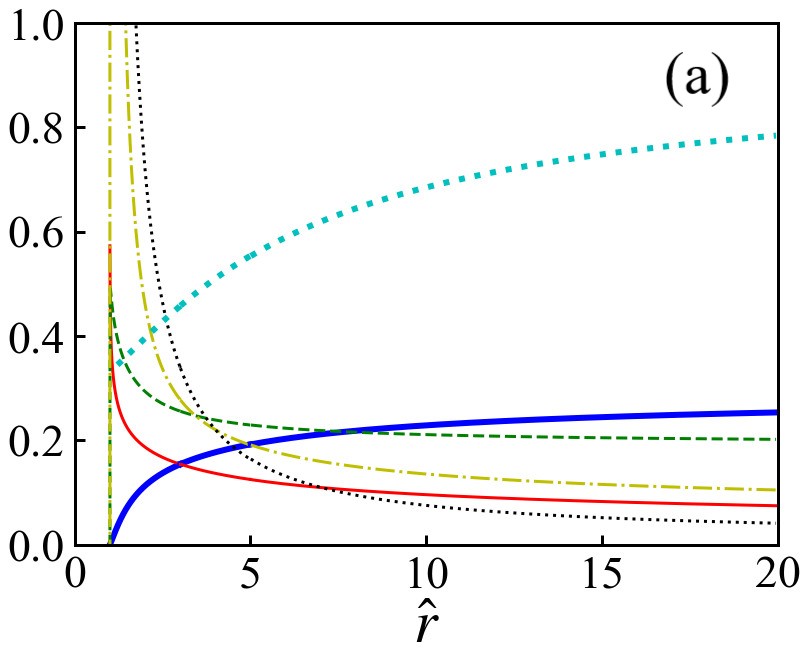}
    \includegraphics[width=80mm,clip]{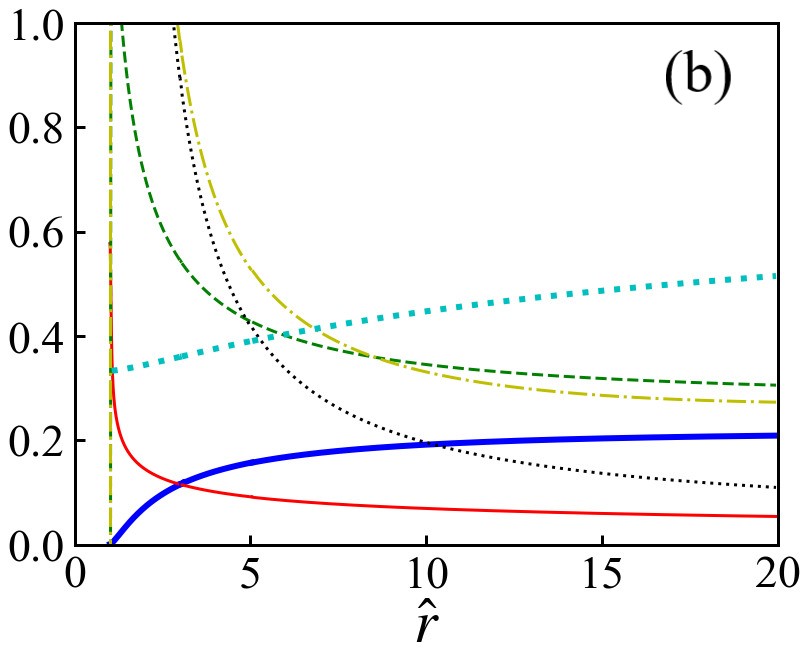}
    \includegraphics[width=80mm,clip]{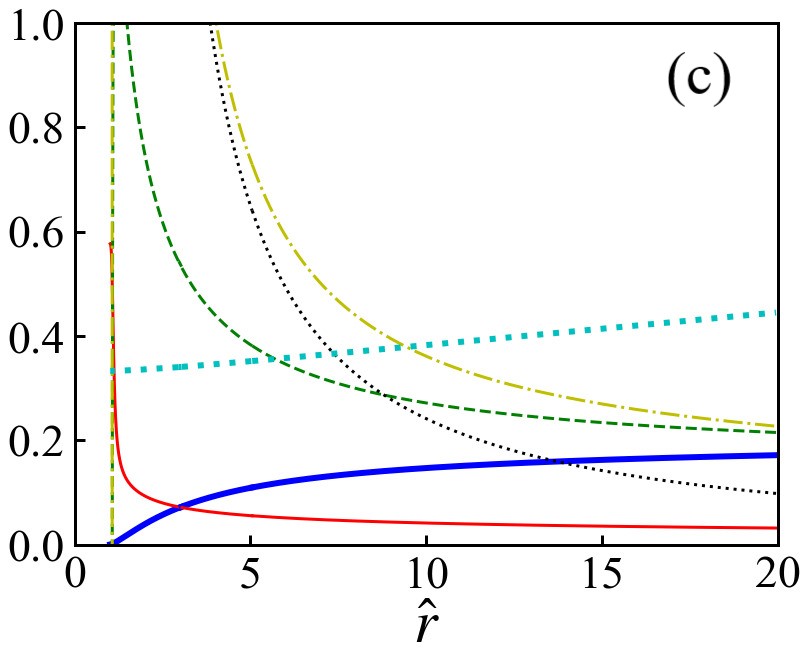}
    \caption{
    Mass-loss rate dependence of transonic solutions.
The critical radius is fixed at $\hat{r}_{\rm c}=3$,
while other parameters are
(a) $\dot{m}=2$, $\dot{e}=3.06$, $m=10^8$, $\Gamma=4/3$, $\hat{L}_{\rm c}=1.28$, $\hat{Q}_{\rm c}=1.37$, $\tau_{\rm c}=1.6$,
(b) $\dot{m}=8$, $\dot{e}=9.574$, $m=10^8$, $\Gamma=4/3$, $\hat{L}_{\rm c}=2.723$, $\hat{Q}_{\rm c}=4.681$, $\tau_{\rm c}=8.9$,
(c) $\dot{m}=20$, $\dot{e}=22$, $m=10^8$, $\Gamma=4/3$, $\hat{L}_{\rm c}=5.37$, $\hat{Q}_{\rm c}=16.49$, $\tau_{\rm c}=32$.
Blue thick solid curves mean $\beta$, red solid ones $\alpha_{\rm s}$, 
green dashed ones $\hat{L}/5$ for (a) and (b), and $\hat{L}/10$ for (c),
yellow chain-dotted ones $\hat{Q}/5$ for (a) and (b), and $\hat{Q}/10$ for (c), 
black dotted ones $\tau/5$ for (a), $\tau/10$ for (b), and $\tau/20$ for (c),
cyan thick dotted ones $f(\beta,\tau)$.
   }
\end{figure}

Fig. 5 shows the mass-loss rate dependence of transonic solutions.
The critical radius is fixed at $\hat{r}_{\rm c}=3$,
while other parameters are
(a) $\dot{m}=2$, $\dot{e}=3.06$, $m=10^8$, $\Gamma=4/3$, $\hat{L}_{\rm c}=1.28$, $\hat{Q}_{\rm c}=1.37$, $\tau_{\rm c}=1.6$,
(b) $\dot{m}=8$, $\dot{e}=9.574$, $m=10^8$, $\Gamma=4/3$, $\hat{L}_{\rm c}=2.723$, $\hat{Q}_{\rm c}=4.681$, $\tau_{\rm c}=8.9$,
(c) $\dot{m}=20$, $\dot{e}=22$, $m=10^8$, $\Gamma=4/3$, $\hat{L}_{\rm c}=5.37$, $\hat{Q}_{\rm c}=16.49$, $\tau_{\rm c}=32$.
Blue thick solid curves mean $\beta$, red solid ones $\alpha_{\rm s}$, 
green dashed ones $\hat{L}/5$ for (a) and (b), and $\hat{L}/10$ for (c),
yellow chain-dotted ones $\hat{Q}/5$ for (a) and (b), and $\hat{Q}/10$ for (c), 
black dotted ones $\tau/5$ for (a), $\tau/10$ for (b), and $\tau/20$ for (c),
cyan thick dotted ones $f(\beta,\tau)$.

As is seen in Fig. 5,
transonic solutions are roughly similar to those in Fig. 4.
In all cases, the gas is quickly accelerated in the vicinity of the black hole
since $\hat{r}_{\rm c}=3$.
In addition, $\hat{L}$ and $\hat{Q}$ take the maximum near the center and vanish at the horizon.
This behaviour is easy understood due to the definitions of $\hat{L}$ and $\hat{Q}$.
As the mass-loss rate $\dot{m}$ increases,
the optical depth also increases as expected.
As the radius increases,
the optical depth decreases,
while the Eddington factor increases.
Near the horizon, on the other hand,
the optical depth diverges,
and therefore, the Eddington factor approaches 1/3.
Near the horizon, furthermore,
$\alpha_{\rm s}$ reaches a relativistic limit of $\alpha_{\rm s}\rightarrow 1/\sqrt{3}$, regardless of the parameters.
In other words, the gas is quite hot near the center, 
and constant $\Gamma$ assumption would be violated, as was stated.
As a result, scattering is dominant in the vicinity of the center,
and absorption would be almost ineffective.

\begin{figure}
    \includegraphics[width=80mm,clip]{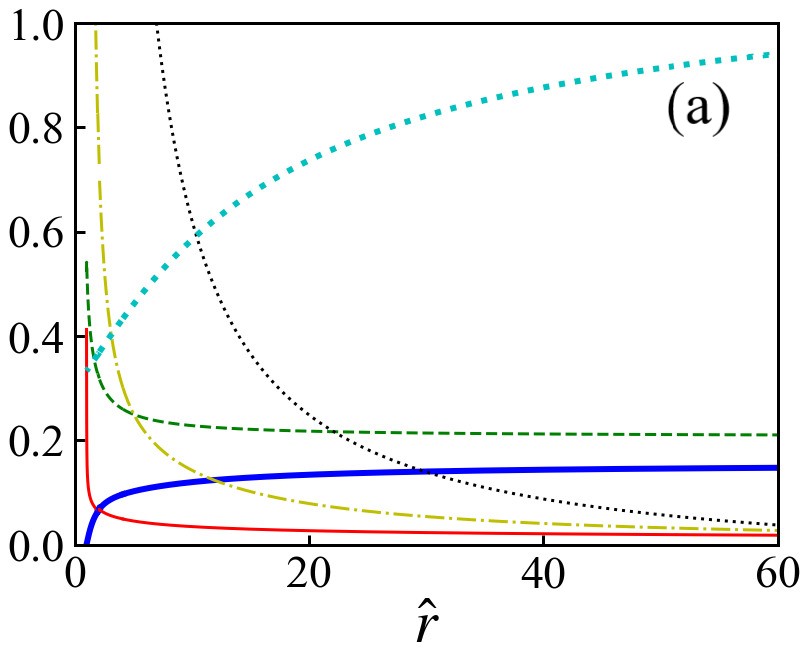}
    \includegraphics[width=80mm,clip]{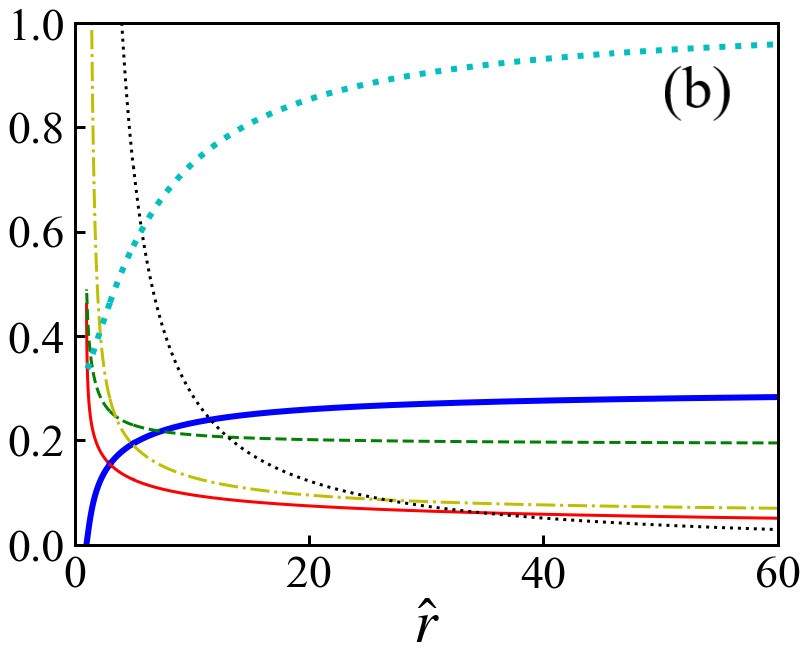}
    \includegraphics[width=80mm,clip]{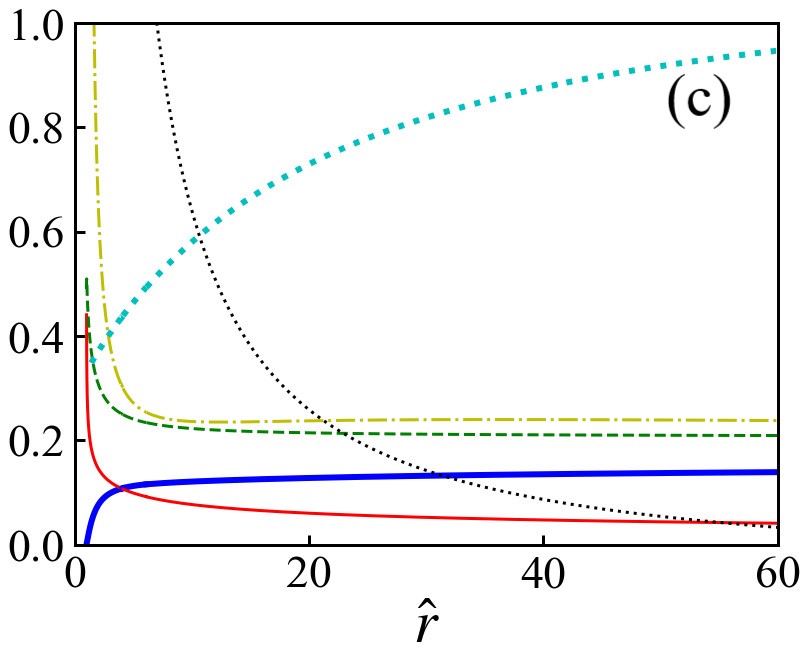}
    \caption{
Critical radius dependence of transonic solutions.
Parameters are fixed as $\dot{m}=2$, $\dot{e}=3.06$, $m=10^8$, $\Gamma=4/3$,
while the critical radius is 
(a) $\hat{r}_{\rm c}=2$ (b) $\hat{r}_{\rm c}=3$, (c) $\hat{r}_{\rm c}=4$.
Blue thick solid curves mean $\beta$, red solid ones $\alpha_{\rm s}$, green dashed ones $\hat{L}/5$,
yellow chain-dotted ones $\hat{Q}/5$, black dotted one $\tau$, cyan thick dotted ones $f(\beta,\tau)$.   
    }
\end{figure}

Fig. 6 shows the critical radius dependence of transonic solutions.
Parameters are fixed as $\dot{m}=2$, $\dot{e}=3.06$, $m=10^8$, $\Gamma=4/3$,
while the critical radius is 
(a) $\hat{r}_{\rm c}=2$ (b) $\hat{r}_{\rm c}=3$, (c) $\hat{r}_{\rm c}=4$.
Blue thick solid curves mean $\beta$, red solid ones $\alpha_{\rm s}$, green dashed ones $\hat{L}/5$,
yellow chain-dotted ones $\hat{Q}/5$, black dotted one $\tau$, cyan thick dotted ones $f(\beta,\tau)$. 

In general, from Bernoulli equation (\ref{bel}),
the terminal velocity $\beta_\infty$ of the winds
increases as $\dot{e}$ increases,
while it decreases as $\dot{m}$ decreases.
However,
the luminosity $\hat{L}_\infty$ also affects the value of the terminal velocity.
Indeed,
the values of $\dot{m}=2$ and $\dot{e}=3.06$ are the same in Fig. 6, 
but there is a slight difference in the velocity and luminosity at the wind top ($\hat{r}=60$); 
for (a) $\beta_{\rm top}\sim 0.148$, $\hat{L}_{\rm top}\sim 1.05$,
for (b) $\beta_{\rm top}\sim 0.283$, $\hat{L}_{\rm top}\sim 0.976$, 
for (c) $\beta_{\rm top}\sim 0.139$, $\hat{L}_{\rm top}\sim 1.05$.
That is,
when $\hat{L}$ is large, $\beta$ becomes small.

\section{Concluding remarks}

We have examined the general relativistic radiatively-driven spherical wind under the nonequilibrium diffusion approximation
with the help of the variable Eddington factor, $f(\tau,\beta)$, 
focusing our attention on the topological nature of critical points,
and application to UFOs.
We found that there appear three types of critical points (loci); saddle, nodal, and spiral types.
The nodal type always admits a deceleration solution, and the spiral type is unphysical. Hence, only the saddle type is reasonable for a transonic accelerated solution.
Furthermore, in order for the terminal speed to be 0.1-0.3$~c$,
the saddle type critical points should be located relatively close to a black hole, as is easily expected.
As a result, the gases are accelerated in the vicinity of the center and pass through the transonic point.
In the nonrelativistic case, it is known that the critical point does not exist for $\Gamma=5/3$.
In the present general relativistic case, however,
the critical point is found to exist even in the $\Gamma=5/3$ case,
due to the relativistic effect,
although there is a non-physical solution near and far from the center that exceeds the relativistic limit of the sound speed.

In the present study, we also bear in mind the applicability of the radiatively driven model for ultra-fast outflows (UFOs).
As was stated, when we calculate transonic solutions, 
we use the parameters ($\dot{e},~\dot{m}$) favorable for UFOs (e.g., Tombesi et al. 2012).
In addition, the luminosity at the wind top was also set to be about the Eddington one.
We thus obtained the transonic solutions, which are consistent with UFOs.
We also found that near the center the gas is so hot that scattering is dominanted and absorption is almost ineffective.
This result is in good agreement with the previous observations, 
and suggests that the radiatively-driven model by continuum is a plausible model
for the acceleration of UFOs. 

In the present paper,
the calculations were made under the simple assumptions of a spherically symmetric
one-dimensional flow, and constant mass-loss rate.
Naturally, the gas cannot be supplied from the black hole itself,
but should be supplied from the very vicinity of the horizon.
One of the main source of the gas is the accretion disc surrounding a black hole, especially, in the present case, the supercritical accretion disc.
If this is the case,
the black hole winds would be realized as funnel jets
(e.g., Lynden-Bell 1978; Fukue 1982; Vyas et al. 2015; Vyas \& Chattopadhyay 2017, 2018).
Another source of the gas could be pair creations
via magnetic fields around the rotating black holes.
However, in the present case
we implicitly assumed the black hole winds
as a baryonic matter.

We further assumed the constant $\Gamma$ in this study for simplicity.
Since the flow temperature becomes quite high in the vicinity of the center,
the value of the ratio of specific heats $\Gamma$ should vary there,
and the temperature dependent $\Gamma$ as well as the relativistic equation of state
should be used there for the precise calculation (cf. Kato et al. 2008).
It is noted that  it may be useful to use the approximation formula of $\Gamma(T)$
and equation of state (e.g., Chattopadhyay \& Ryu 2009). 

As was stated, 
we expected the gas to be sufficiently hot near the center 
and assumed that the gas is fully ionized. 
As a result, we ignored the line-driven force.
However, at the outer region where the flow temperature decreases,
the line-driven mechanism could work.
If the line-force dominates the continuum one,
the flow could be further accelerated in the outer region.

These considerations are for the future works.


\section*{Acknowledgements}

The author would like to thank an anonymous referee
for valuable comments,
which improved the original manuscript.

This work has been supported in part
by a Grant-in-Aid for Scientific Research (18K03701) 
of the Ministry of Education, Culture, Sports, Science and Technology.

\section*{Data availability}

No new data were generated or analysed in support of this research.

\bsp

\label{lastpage}

\end{document}